\documentclass[pdflatex,sn-mathphys]{sn-jnl}
\usepackage{times}
\usepackage{amsmath}
\usepackage{graphicx}
\usepackage{color}
\usepackage{comment}
\usepackage[displaymath,mathlines]{lineno}
\newcommand {\red}[1]{{#1}}  

\begin{document}

\title[Quantum harmonic free energies]{Quantum harmonic free energies for biomolecules and nanomaterials}

\author[1,2]{\fnm{Alec F.} \sur{White}}

\author[1]{\fnm{Chenghan} \sur{Li}}

\author[1]{\fnm{Xing} \sur{Zhang}}

\author[1]{\fnm{Garnet Kin-Lic} \sur{Chan}}

\affil[1]{\orgdiv{Division of Chemistry and Chemical Engineering}, \orgname{California Institute of Technology}, \orgaddress{\street{1200 East California Boulevard}, \city{Pasadena}, \postcode{91125}, \state{CA}, \country{United States of America}}}
\affil[2]{Present address: Quantum Simulation Technologies, Inc., Cambridge, MA 02139, \country{United States of America}}

\date{June 2021}

\abstract{
Obtaining the free energy of large molecules from quantum mechanical energy functions is a longstanding challenge. We describe a method that allows us to estimate, at the quantum mechanical level, the harmonic contributions to the thermodynamics of molecular systems of large size, with modest cost. Using this approach, we compute the vibrational thermodynamics of a series of diamond nanocrystals, and show that the error per atom decreases with system size in the limit of large systems. We further show that we can obtain the vibrational contributions to the binding free energies of prototypical protein-ligand complexes where the exact computation is too expensive to be practical. Our work raises the possibility of routine quantum mechanical estimates of thermodynamic quantities in complex systems.
}

\maketitle


The contributions to the free energy from atomic motion are critically important to the thermodynamics and kinetics of biological, chemical, and materials systems. Changes in such contributions govern processes ranging from the affinity of drug binding to structural phase transitions in crystals.
When the internal energy is computed at the quantum mechanical level, 
a harmonic approximation is often the only feasible option to describe atomic motion. 
However, for large systems such as nanostructures and biomolecules, computing free energy contributions is expensive even within the harmonic approximation. For a system of $N$ atoms, the Hessian matrix which describes the vibrations requires $O(3N)$ gradient calculations, or $O(3N)$ times the cost of computing the internal energy. This is clearly prohibitive when $N$ is large, making free-energy computation in large systems with quantum mechanical methods a major contemporary challenge~\cite{grimme2018computational}.

There are many possible strategies to speed up harmonic vibrational analysis, including methods based on a partial Hessian\cite{Li2002,Woodcock2008}, iterative diagonalization \cite{Filippone2001}, and Hessian-free methods that use molecular dynamics to approximate the the harmonic problem\cite{Karplus1981,Brooks1995}. Here we describe a different strategy where we estimate vibrational thermodynamic quantities directly without ever computing the full Hessian or taking advantage of any local structure. 


The starting point is to express each harmonic thermodynamic quantity as a matrix function trace.
Then, our technique contains three  elements. 
First, we sample the matrix trace operation using random vectors and stochastic Lanczos quadrature~\cite{Ubaru2017}. Second, we compute the Hessian-vector product at the same cost as the gradient from the difference of gradients at displaced geometries, bypassing the  Hessian construction entirely. Third, we ameliorate the stochastic error, especially for free energy differences, through a form of correlated sampling.
Related stochastic methods have been used for anharmonic corrections to the harmonic free energy\cite{Hellman2013,Errea2014}, as well as in stochastic electronic structure\cite{Baer2013}, but to our knowledge this is the first time these ideas have been brought to bear on the harmonic thermodynamic quantities themselves. As we demonstrate, this  allows us to compute at the quantum mechanical level and with modest cost, vibrational free energy contributions for nanocrystals with more than 600 atoms, and free energy differences in protein-ligand complexes with more than 3000 atoms.

\section*{Theory}

We first express the harmonic thermochemical quantities as traces of matrix functions.  In particular, we are interested in the zero-point energy (ZPE),
\begin{linenomath*}
\begin{equation}
    \mathrm{ZPE} = \sum_I \frac{\omega_I}{2} = \sum_I\frac{\sqrt{\omega_I^2}}{2} = \mathrm{Tr}\left[ \frac{\sqrt{\bf{D}}}{2} \right]
\end{equation}
\end{linenomath*}
the thermal contribution to the enthalpy,
\begin{linenomath*}
\begin{equation}
	H_{\mathrm{vib}} - \mathrm{ZPE} = \sum_I \omega_I\left(\frac{e^{-\beta\omega_I}}{1 - e^{-\beta\omega_I}}
	\right) = \mathrm{Tr}\left[\frac{\sqrt{\bf{D}}\exp(-\beta\sqrt{\bf{D}})}{1 - \exp(-\beta\sqrt{\bf{D}})}\right]
\end{equation}
\end{linenomath*}
and the entropy,
\begin{linenomath*}
\begin{equation}
\begin{split}
	S_{\mathrm{vib}}/k_B  &= \sum_I \left[\beta\omega_I\frac{e^{-\beta\omega_I}}{1 - e^{-\beta\omega_I}} -\ln\left(1 - e^{-\beta\omega_I}\right) \right] \\
	&= \mathrm{Tr}\left[\beta\frac{\sqrt{\bf{D}}\exp(-\beta\sqrt{\bf{D}})}{1 - \exp(-\beta\sqrt{\bf{D}})} - \ln\left(1 - \exp(-\beta\sqrt{\bf{D}})\right)\right]
\end{split}
\end{equation}
\end{linenomath*}
Here $\beta$ is the inverse temperature, $\{\omega_I\}$ is the set of normal mode frequencies, and $\bf{D}$ is the mass-weighted Hessian matrix. We refer to ZPE as a non-thermal quantity as it has no temperature dependence.

The above expressions have the form $\mathrm{Tr} f(\mathbf{D})$. We now employ a stochastic estimator of the trace. The simplest version writes $\mathrm{Tr} f(\mathbf{D}) \approx \frac{M}{n} \sum_{l=1}^{n} \mathbf{v}_l^\mathrm{T} f(\mathbf{D}) \mathbf{v}_l$ where $\mathbf{v}_l$ are a set of $n$ random vectors with zero mean and unit covariance, and $M=3N-6$ is the dimension of $\mathbf{D}$. This direct stochastic evaluation requires a polynomial approximation of $f(\mathbf{D})$, which is typically carried out using a Chebyshev expansion~\cite{han2015large}. A closely related idea, which we use in this work, is stochastic Lanczos quadrature\cite{Ubaru2017}. In this technique, the polynomial approximation is generated for the scalar  $\mathbf{v}^T_l f(\mathbf{D}) \mathbf{v}_l$ rather than globally for the function $f(\mathbf{D})$. We have found the stochastic Lanczos method to be slightly superior to the Chebyshev polynomial approach for the quantities in this work.
Within the polynomial expansion, the main operation is the matrix-vector product $\mathbf{D} \mathbf{x}$, where $\mathbf{x}$ is in the stochastic Lanczos space. This can be computed from the difference of gradients, displaced by $\delta \mathbf{x}$ in mass-weighted coordinates, for small $\delta$~\cite{kaledin2005gradient}. Thus no Hessian is needed at all in this approach (further details are provided in the Methods section). We note the sampling itself does not depend on $f$, and thus intermediate information, such as Lanczos quadrature weights and positions can be cached to estimate $\mathrm{Tr}f(\mathbf{D})$ for any other $f$, and as such thermostatistical quantities at different temperatures can be computed without the need for repeating the sampling procedure.

Within the above scheme, there are two sources of error. The first is from the order of the Lanczos quadrature, $m$. This vanishes when $m$ is greater than or equal to the matrix dimension.
The second is the sampling error, which decreases like $1/\sqrt{n}$ for $n$ random  vectors. 
For a quadrature order of $m$ and $n$ samples, the cost of the method is equal to $O(mn)$ gradient calculations. To reduce the statistical error, we use a form of correlated sampling. For the absolute thermodynamic quantities computed for the diamond nanocrystals we employ a high-level quantum mechanical approach as well as a cheaper low-level method (for example, a force-field, or semi-empirical quantum-mechanical approach) where the exact computation of the harmonic thermodynamic quantity $X_\text{low}$ is possible. Then, the free energy contribution for the high-level method is obtained as
\begin{linenomath*}
\begin{equation}
    X_\text{high} = X_\text{low} + \Delta
\end{equation}
\end{linenomath*}
where $\Delta$ is computed by applying the stochastic Lanczos quadrature to the difference of high-level and low-level methods. 
In the case of protein(P)-ligand(L) binding, we are interested in the difference between the holo (ligand-bound) state and the apo (ligand-free) state, i.e.
\begin{linenomath*}
\begin{equation}
X_\text{bind} = X_\text{P+L} - X_\text{P} - X_\text{L}
\end{equation}
\end{linenomath*}
where $X_\text{bind}$ represents the ligand binding free energy, enthalpy, entropy, etc. In this case, we perform correlated sampling by using the same random vectors in the stochastic Lanczos treatment of the P+L, P, L systems (zeroing out elements for P and L respectively). No additional low-level method is involved in ligand-binding calculations.

\section*{Results}

\begin{figure}
    \centering
    \includegraphics[scale=0.75]{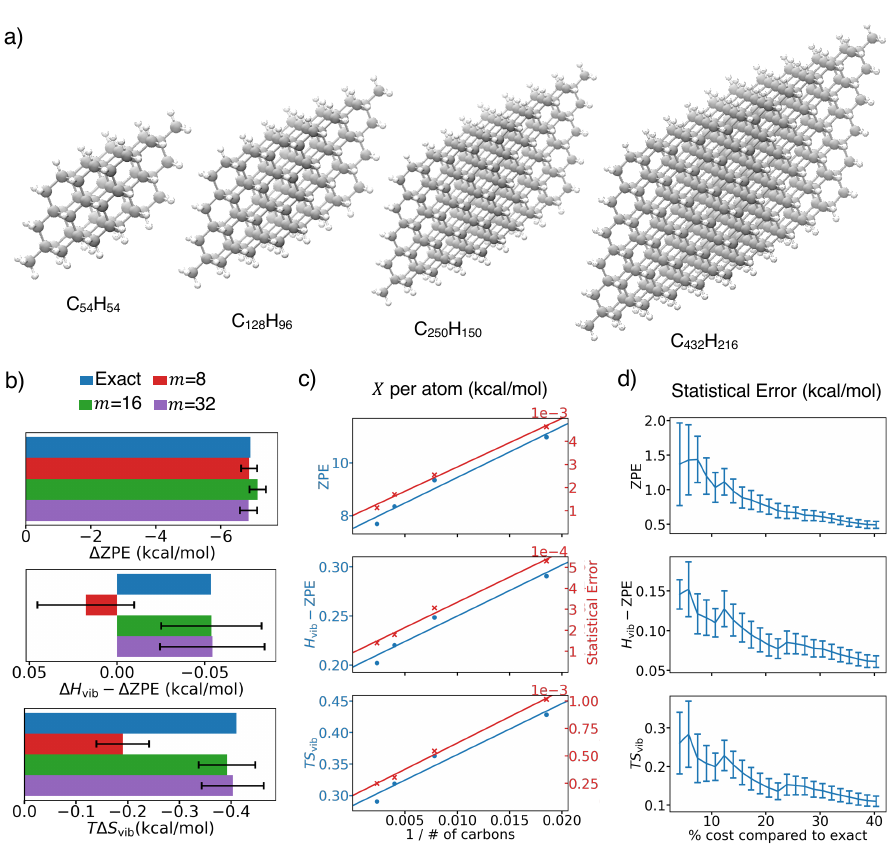}
    \caption{ 
    {\bf a) } Structures and chemical formulae of the diamond nanocrystals used in the calculations.  
    {\bf b) } $\Delta$ values estimated with different Lanczos quadrature orders ($m$) for the smallest system ($\text{C}_{54}\text{H}_{54}$) compared to the exact values. Error bars were estimated using the standard error of 50 samples.
    {\bf c)} Per-atom quantity values and errors as a function of system size for a fixed number of samples (50 samples). The solid lines correspond to linear fits to inverse size. Note that the scale used for the quantities and their errors is different for each of the 3 graphs.
    {\bf d) } Statistical errors (for $m=16$ quadrature) in absolute thermodynamic quantities of $\text{C}_\text{432}\text{H}_\text{216}$ as a function of \% of computational cost of the exact calculation (error bars denote error of error).
    }
    \label{fig:NP1}
\end{figure}

As a first application, we take diamond nanocrystals (Figure~\ref{fig:NP1}a) as prototypical nanomaterials, and compute the free energies as a function of size.
We employ the high- and low-level correlated sampling approach described above, with Kohn-Sham density functional theory (DFT) with the PBE functional~\cite{perdew1996generalized} as the high-level method and the semi-empirical extended tight-binding (xTB) method~\cite{grimme2017} as the low-level method.
For the smallest system ($\mathrm{C_{54}H_{54}}$)
, we can compute the Hessian explicitly to provide an exact reference.  
Figure~\ref{fig:NP1}b shows $\Delta$ quantities for the zero-point energy, the thermal enthalpy, and the entropy respectively, as a function of quadrature order $m$. 
The error bar indicates the statistical error for 50 samples (see SI for details of error analysis). 
A stochastic quadrature level of $m = 8$ does not provide sufficient accuracy, so we choose $m = 16$ for further calculations. Additional calculations on three transition-metal complexes using $m=16$ are presented in the SI, although the choice of $m$ in general is system and accuracy specific.
Figure~\ref{fig:NP1}c shows the value and stochastic error per atom (estimated as one standard error) for the ZPE, thermal enthalpy, and entropy respectively. We note the error decreases with the size of the system, faster than the decay of the quantities themselves, which is evidence of ``self-averaging'' due to the large system size. \red{A more detailed discussion of the ``self-averaging'' behavior is provided in the SI. }
Thus if one is interested in per-atom quantities, as is often the case for thermodynamics, for example to locate phase transitions, our stochastic approach becomes increasingly more efficient in a large system. The difference between our largest simulation and the extrapolated thermodynamic limits for the per-atom ZPE, enthalpy, and entropy is only 0.2 kcal/mol, 0.004 kcal/mol, and 0.006 kcal/mol respectively; statistical errors with 50 samples are about 0.001 kcal/mol or less. In fact, in the largest diamond system,
with a \emph{single} sample, one can estimate the per-atom quantities with a statistical error of less than 0.01 kcal/mol (this error was estimated from 50 samples; see SI for details) at a 120-fold speedup relative to the exact Hessian calculation.


If one is instead interested in the absolute values, the method can still be cheaper than the full computation of the Hessian. 
Figure~\ref{fig:NP1}d shows the stochastic error for the largest carbon system ($\mathrm{C_{432}H_{216}}$) as a function of computational cost. At less than 20\% of the exact calculation's computational effort, the error estimate for ZPE is well within 1 kcal/mol, red corresponding to 0.03\% relative error with respect to the total ZPE or 7.7\% relative error with respect to the $\Delta$ZPE between DFT and xTB. Less precise estimates can be obtained even more cheaply; a 20 times speedup is possible if 2 kcal/mol of error in the absolute quantities is tolerable. 

Vibrational contributions to the free energy are also central to the study of large molecule and biomolecular interactions. For protein-ligand interactions in particular, where the aggregate binding is often only 5-15 kcal/mol, the vibrational contribution to binding free energies can be significant. Although the harmonic approximation is not necessarily a faithful approximation in these systems, the harmonic contributions nonetheless provide a useful first estimate of the thermal and entropic contributions~\cite{ehrlich2017towards, spicher2020efficient}. The task is challenging for the stochastic Lanczos approach as binding is the difference of large absolute quantities, requiring tight convergence of the statistical error.
To reduce statistical error, 
we use the fixed random vector correlated sampling approach described above.
We do not include explicit water molecules in the simulation: in principle, these could be included at additional cost, or the desolvation contribution to the free energy can be separately estimated by standard continuum methods~\cite{ehrlich2017towards}.

\begin{figure}
    \centering
    \includegraphics[scale=0.75]{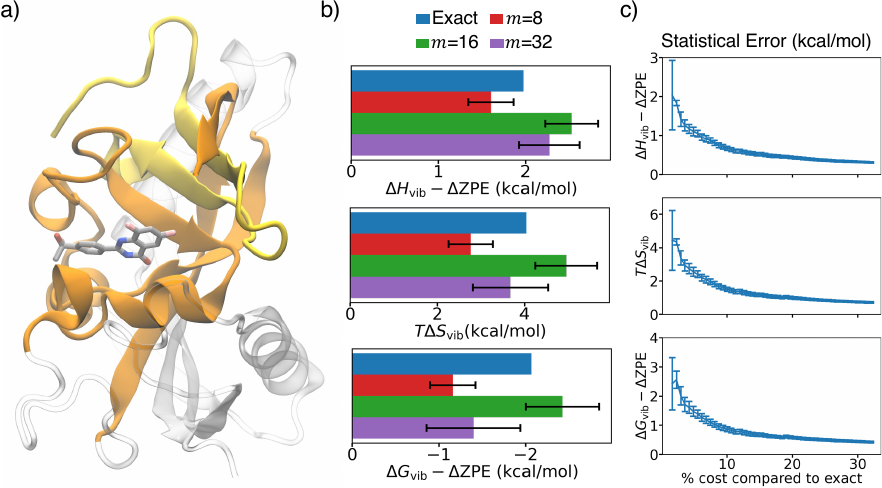}
    \caption{
    Harmonic thermal contributions for the TNKS2 complex computed at the xTB level.
    {\bf  a)} An image of the TNKS2 complex. The truncated part of the protein is shown as transparent, while the remaining two protein chains and the ligand are colored orange, yellow and grey, respectively. Image rendered by VMD\cite{vmd}. 
    {\bf b)} The thermal enthalpy and entropy, and free energy of binding for the TNKS2 system for varying Lanczos order. The error bars represent $\pm$ one standard error from 100 random samples. 
    {\bf c)} Statistical errors in binding free energy quantities as a function of \% of cost of the exact calculation (error bars denote error of error).
    }
    \label{fig:PL15}
\end{figure}

We first study a system which is just small enough that exact results at the xTB level can be obtained at a high computational cost: a cutout ($\sim$ 1600 atoms) of the human tankyrase 2 (TNKS2) protein with a bound ligand shown in Figure~\ref{fig:PL15}a (see Methods for more information). 
In Figure~\ref{fig:PL15}b we show the thermal contributions to the binding enthalpy, entropy and free energy for the stochastic Lanczos quadrature orders $m=8, 16, 32$ using xTB. \red{We present a detailed check of the Lanczos convergence for a larger set of $m$ values, and across a set of different systems, in the SI (see Figures 3S and 4S). Following these convergence checks,} we choose $m=16$ for further calculations, where
the error due to the Lanczos order is estimated to be less than 1 kcal/mol.
 Table \ref{tab:HIV} 
 summarizes the data using up to 100 samples for all rovibrational free energy contributions (the rotational contribution is obtained following Ref.~\cite{grimme2012supramolecular}). From comparison to the exact results, the total thermal contribution $\Delta G_\text{vib}-\Delta \text{ZPE}$ can be estimated with a statistical error of less than 1 kcal/mol with a cost of roughly 10\% of the exact Hessian calculation. The non-thermal contribution $\Delta \text{ZPE}$ (not plotted) has larger statistical error, but can still be estimated to better than 2 kcal/mol with roughly 200 samples, or 67\% cost of the exact Hessian calculation.

We next evaluate the thermal quantities at the Kohn-Sham DFT level using the PBE functional for the TNKS2 complex using up to 35 samples,  as summarized in Table~\ref{tab:HIV}. Interestingly, we find the thermal vibrational contributions at the DFT level to be quite similar to those from xTB. 
A single sample using our DFT implementation takes roughly \red{two days} on 1 node (32 CPU cores), compared to 600 days on 1 node for the exact Hessian calculation.

\begin{figure}
    \centering
    \includegraphics[scale=0.8]{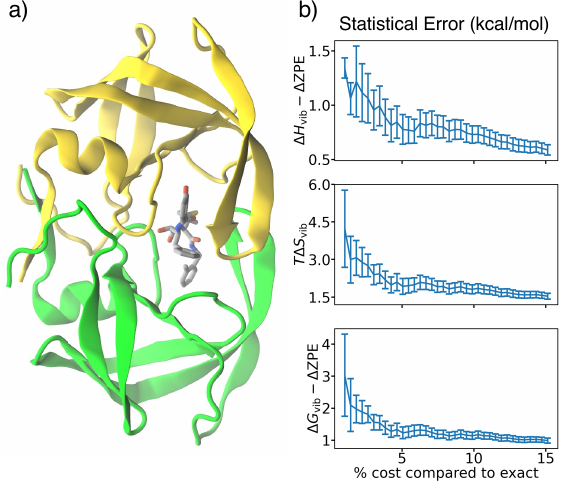}
    \caption{
    Harmonic thermal contributions for the JE-2147-HIV protease complex computed at the xTB level.
    {\bf a)} An image of the JE-2147-HIV protease complex. The two protein chains and the ligand are shown in yellow, green, and grey respectively.  Image rendered by VMD\cite{vmd}.
    {\bf b)} Statistical errors in binding free energy quantities as a function of \% of cost of the exact calculation (error bars denote error of error).
    }
    \label{fig:1kzk}
\end{figure}

Finally, we apply our approach to 
HIV protease bound to a small molecule (JE-2147)\cite{Reiling2002} (Figure~\ref{fig:1kzk}a). This system contains over 3000 atoms including hydrogens. 
The number of gradient calculations required to compute the full Hessian in this case ($\sim$ 10,000) is so large that the exact computation is expensive
even at the level of semi-empirical quantum mechanics, thus we do not compute exact data here.
\begin{table}
    \centering
    \begin{tabular}{c|ccccc}
        \hline\hline
         Quantity & TNKS (xTB) & TNKS (DFT) & HIV (xTB) & TNKS (expt.) & HIV (expt.) \\
         \hline
         $\Delta G_\mathrm{rot}$ & 9.57 & 9.61 & 10.43 & & \\
         $\Delta H_{\mathrm{vib}}-\Delta$ZPE & 2.53$\pm$0.30  & 2.89$\pm$0.51 &  1.88 $\pm$ 0.72 & & \\
         $T\Delta S_{\mathrm{vib}}$ &          4.96$\pm$0.71  & 5.79$\pm$1.11     &    1.69 $\pm$ 1.79 & & \\
         $\Delta G_{\mathrm{vib}}-\Delta$ZPE & -2.43$\pm$0.42  &-2.90$\pm$0.64      &  0.19 $\pm$ 1.13 & & \\
         $\Delta G_\mathrm{bind}^\mathrm{tot}$ & -26.8 & -21.4 & -14.1 & -11.0 & -14.2  \\
         \hline\hline
    \end{tabular}
    \caption{Contributions to the binding free energy at 298.15 K for the TNKS2 and JE-2147-HIV protease system. The rotational free energies $\Delta G_\text{rot}$ are computed at the optimized structures, the thermal enthalpy, entropy and free energy are computed from stochastic sampling. The error estimate for TNKS using xTB is one standard error from 100 random samples (corresponding to 33\% of exact cost), 35 samples (corresponding to 11\% of exact cost) for TNKS using DFT, and 50 random samples (corresponding to 10\% of exact cost) for HIV (xTB). 
    \red{The experimental binding affinity is estimated as either $k_B T \ln{\mathrm{IC}_{50}}$ or $k_B T \ln{K_i}$, with IC$_{50}=$ 9.2 nM for TNKS2 from Ref.~\cite{buchstaller2019discovery} and $K_i=$ 41 pM for HIV protease from Ref.~\cite{Reiling2002}. $\Delta G_\mathrm{bind}^\mathrm{tot}$ includes the non-thermal contribution from the ZPE and is computed as described in the Methods section.} All values are given in kcal/mol. }
    \label{tab:HIV}
\end{table}
In Table~\ref{tab:HIV} we show our stochastic estimates of the harmonic contributions to the thermodynamic binding quantities using xTB and a quadrature order of $m = 20$. 
Interestingly, the thermal contributions ($\Delta H_\text{vib}-\Delta$ZPE and $T\Delta S_\text{vib}$) to the free energy are both similar, small, and of opposite sign, meaning that the total thermal free energy contribution is almost zero. \red{The small size of the thermo-statistical harmonic contributions may be due to the known rigidity of the HIV protease binding pocket, which means that many of the normal modes in the free protein and protein-ligand complex may be very similar.} Nonetheless, estimating $\Delta G_\text{vib}-$ZPE to an accuracy of 1 kcal/mol is clearly feasible within our scheme at roughly 10\% of the estimated cost of the exact calculation (Figure~\ref{fig:1kzk}b). 

\begin{table}[]
    \centering
    \begin{tabular}{c|cccc}
    \hline \hline
  Quantity & All Diamonds & TNKS (xTB) & TNKS (DFT) & HIV (xTB)\\ 
  \hline
$\Delta H_{\mathrm{vib}}-\Delta$ZPE & $<$1\% &  3\% &  3\% & 11\% \\
$T\Delta S_{\mathrm{vib}}$          & $<$1\% & 16\% & 14\% & 74\% \\
$\Delta G_{\mathrm{vib}}-\Delta$ZPE & $<$1\% &  6\% &  5\% & 31\% \\
\hline\hline
    \end{tabular}
    \caption{Summary of computational cost to achieve an accuracy of 1 kcal/mol compared to an exact Hessian calculation. The number of samples $n$ required to achieve the desired accuracy is estimated with 50 samples for the diamond systems ($n'=50$), $n'=100$ for TNKS2 (xTB), $n'=35$ for TNKS (DFT), and $n'=50$ for HIV (xTB). See SI for details about error analysis. }
    \label{tab:speedup}
\end{table}

\section*{Discussion}

We have presented results which demonstrate the feasibility of computing harmonic contributions to the free energy at the quantum mechanical level for systems of more than a thousand atoms. The cost is greatly reduced from that needed to compute the Hessian of the system. This is particularly true when one is interested in intensive (or ``per-atom'') quantities, where self-averaging behavior shows that in large systems, we may estimate the quantities at a cost comparable to that of a few energy evaluations. This holds promise in evaluating thermodynamic transitions in materials involving large unit cells, for example, those associated with alloys and disorder. In the case of free energy differences, the correlated sampling technique employed here makes the evaluation of even small thermal free energy differences, as found in protein-ligand complexes, feasible at the level of 1-2 kcal/mol. The estimated speedups for all systems considered, in order to reach a given accuracy, are summarized in Table~\ref{tab:speedup}.

An additional advantage of the current approach is that the cost may be continually tuned. This is relevant to new applications, for example in the computational screening for therapeutics~\cite{grimme2018computational,mardirossian2020novel}, where less precise estimates are an acceptable tradeoff for speed. 
\red{Also, as we see from Table~\ref{tab:HIV}, while the computed binding affinity using the harmonic approximation is not always  highly accurate in biomolecular systems}, the increased facility to obtain harmonic estimates further raises the possibility for new approaches to compute anharmonic contributions to free energies with a variety of techniques, \red{such as the minimum-mining technique~\cite{chen2010modeling}, which samples multiple minima in an anharmonic potential and combines harmonic contributions from each of them.} While applications which require intensive quantities benefit most from the stochastic approach, converging absolute thermal quantities to sufficiently high precision may require improved statistical estimators~\cite{meyer2021hutch++}. In addition, while we have estimated harmonic thermal contributions using quantum mechanical energy functions, the same algorithm accelerates harmonic free energy computation using any energy function, including classical force-fields, and can be combined with other cost reduction techniques, such as partial Hessians.

In summary, the technique presented here suggests that the estimation of harmonic free energy effects at the quantum mechanical level for systems with hundreds or even more than a thousand of atoms need not be considered a future challenge~\cite{grimme2018computational}, but one which can be begin to be addressed today.

\section*{Methods}

\subsection*{Stochastic Lanczos quadrature}

The stochastic Lanczos method is a numerical method which has been employed   in different contexts (see e.g. Ref.~\cite{jaklivc1994lanczos} for an early application in quantum many-body systems). We follow the general mathematical formulation in  Ref.~\cite{Ubaru2017}. The Lanczos iterations were performed starting from a vector randomly selected from a Rademacher distribution. The Lanczos iterations require the action of the mass-weighted Hessian matrix on this random vector. We compute this matrix-vector product from finite difference gradient calculations:
\begin{linenomath*}
\begin{equation}
    \mathbf{D}\mathbf{v} = \mathbf{M}^{-1/2}\frac{\mathbf{g}(\delta\mathbf{v}) - \mathbf{g}(-\delta\mathbf{v})}{2\delta}.
\end{equation}
\end{linenomath*}
Here, $\bf{D}$ is the mass-weighted Hessian matrix, $\mathbf{g}$ is the gradient, $\bf{M}$ is the diagonal matrix of masses. The displacement is given by
\begin{linenomath*}
\begin{equation}
    \delta \mathbf{v} = \mathbf{M}^{-1/2}\mathbf{v}
\end{equation}
\end{linenomath*}
where the factor of $\mathbf{M}^{-1/2}$ accounts for the mass-weighting. The value of $\delta$ is chosen based on the norm of the random vector so that the average displacement per atom is 0.0012 \AA.

The number of Lanczos iterations $m$ is a parameter of the method; $m$ should be increased until convergence is reached. In chemical systems, the maximum eigenvalue of the Hessian does not scale with the system size (it is the maximum vibrational frequency, for example a C-H stretch), while the minimum eigenvalue is bounded from below by $0$. For many functions of the Hessian, this means that the maximum and minimum eigenvalues also do not scale with system size. Under this assumption, we expect a fixed $m$ to yield a constant relative error in the trace of the function, and furthermore, due to exponential convergence in $m$ for well-behaved functions of the Hessian, $m$ needs to only increase logarithmically with system size for constant absolute error. A numerical study of convergence with $m$ is presented in the SI.

Additionally, we also implemented and tested a Chebyshev fitting method as an alternative to the stochastic Lanczos quadrature. We found that often a higher order Chebyshev fit was required making it a slightly more expensive alternative to Lanczos quadrature. 

\subsection*{Calculations on diamond nanocrystals}
Diamond nanocrystals were constructed by creating supercells of the bulk diamond unit cell and then capping with hydrogens. The resulting structure was optimized using the PBE functional\cite{perdew1996generalized} and the def2-SV(P) basis set\cite{weigend2005balanced}. All DFT calculations were performed with the ORCA program\cite{Neese2012, Neese2018}. The structure was also optimized using the second generation extended tight-binding (GFN2-xTB) method\cite{Bannwarth2019} as implemented in the Semiempirical Extended Tight-Binding (xTB) program package\cite{Bannwarth2021}. 

\subsection*{Calculations on protein-ligand systems}
All calculations on protein-ligand systems used the second generation extended tight-binding (GFN2-xTB) method\cite{Bannwarth2019} as implemented in the Semiempirical Extended Tight-Binding (xTB) program package\cite{Bannwarth2021}. Generalized Born, solvent-accesible area (GBSA) solvation was used to mimic an aqueous environment for all calculations. For the TNKS2 system, we additionally performed the calculations using density functional theory. We used the PBE functional with the GTH-DZV basis and GTH pseudo-potential\cite{hartwigsen1998relativistic} for PBE. The system was placed in a $\mathrm{45.55\AA\times41.78\AA\times37.27\AA}$ periodic box, allowing a 5-6$\mathrm{\AA}$ vacuum around the atoms. A 0.15 Hartree level shift was applied to the virtual orbitals to help the SCF convergence. The Gaussian and Plane Waves method\cite{lippert1997hybrid, mcclain2017gaussian} was employed and the plane wave cutoff was 200 Hartree. 

The truncated TNKS2 protein was constructed from the the ligands/protein-structure obtained from Ref.~\cite{Schindler2020}. The entire protein was minimized using AmberTools, using the Generalized Born implicit, igb=5), the Amber 14 force field\cite{Maier2015}, and the general AMBER force field (GAFF)\cite{Wang2004} for ligands, assigned using Antechamber from AmberTools\cite{Case2021}. Following minimization, truncation and capping of the terminals were carried out using PyMol~\cite{pymol}. Truncation was performed to remove all protein atoms beyond ~3-4Å around the ligand. Truncated ends were capped using ACE/NME terminal patches. The ligand bound to the protein is one of the many inhibitors identified in Ref.~\cite{Waaler2020} whose structure is available in the Protein Data Bank \cite{Berman2002}(PDB: JKN). 

The structure of the JE-2147-HIV protease complex was obtained from PDB 1KZK. Hydrogens were added using UCSF Chimera\cite{Pettersen2004} and the structure was optimized first using the GFN-FF force field as implemented in the xTB program package\cite{Bannwarth2021} and finally with the GFN2-xTB method\cite{Bannwarth2019} ultimately used for harmonic vibrational analysis.

\red{The total binding free energy is estimated as $\Delta G_\mathrm{rot} + \Delta H_\mathrm{vib} -T\Delta S_\mathrm{vib} + \Delta E + \Delta G_\mathrm{solv}$, where $\Delta E$ is the single point energy difference and $\Delta G_\mathrm{solv}$ is estimated via GBSA within xTB.}

\backmatter

\bmhead{Supplementary information}
Supplementary information on statistical analysis, additional analysis on three transition metal complexes. 

\bmhead{Acknowledgments}
We thank S. Murlidaran for help with protein preparation.

\section*{Declarations}

\begin{itemize}
\item Funding. Work by Alec F. White was supported by the US Department of Energy, via grant no. DE-SC0018140. Work by Chenghan Li was supported by the US National Science Foundation via grant no. 1931328. 

\item GKC is a part owner of QSimulate, Inc.  
\item Availability of data and materials: Data is available from the authors upon reasonable request.
\item Authors' contributions: AFW and GKC conceived the project. AFW, CL carried out the work. All authors contributed to the writing of the paper.
\end{itemize}

\bibliography{ref}

\end{document}


\maketitle

\section{Error estimation}
For each random vector $\mathbf{v}_l$, $s_l = M \times \mathbf{v}_l^\mathrm{T} f(\mathbf{D}) \mathbf{v}_l$ forms an unbiased estimator of $\mathrm{Tr}(f(\mathbf{D}))$, and so does the average of $n$ samples, $\Bar{s}_n=\frac{1}{n}\sum_l^n s_l$. Since each random vector is drawn independently, it is well-known that the sample variance $S^2$ (defined below) is an unbiased estimator of Var($\Bar{s}_n$), i.e., $ \mathbb{E}[(\Bar{s}_n-\mathbb{E}[\Bar{s}_n])^2]$, and the stochastic error in the $n$-sample estimator $\Bar{s}_n$ can be estimated as follows 
\begin{equation}
    S_n=\sqrt{\frac{1}{n(n-1)} \sum_l^n (s_l-\Bar{s}_n)^2}\tag{1S}
    \label{eq:S}
\end{equation}
Such an $S_n$ is sometimes referred to as the standard error, and we follow this convention in our main text. To estimate the error of $S_n$, we would like to estimate $\mathrm{Var}(S_n^2)=\frac{1}{n} (\mu_4 - \frac{n-3}{n-1}\mu_2^2)$ from our finite samples, where $\mu_p=\mathbb{E}[(s_l-\mathbb{E}[s_l])^p]$. We use unbiased estimators for $\mu_4$ and $\mu_2^2$ (sometimes known as (generalized) $h$-statistics)
\begin{equation}
    h_4 = \frac{(n^3-2n^2+3n)m_4-n(6n-9)m_2^2}{(n-1)(n-2)(n-3)} \tag{2S}
\end{equation}
\begin{equation}
    h_{2,2} = \frac{-(n^2-n)m_4+n(n^2-3n+3)m_2^2}{(n-1)(n-2)(n-3)} \tag{3S}
\end{equation}
where $m_p$ is the $p$-th central moment of the samples, such that $\frac{1}{n} (h_4 - \frac{n-3}{n-1}h_{2,2})$ forms an unbiased estimator for Var$(S_n^2)$, and the error of error is computed from its square root.

We next consider error estimation for $\Bar{s}_n$ from $n'$ samples where $n'>n$. A special case is to estimate the error of a single-sample estimator $\Bar{s}_1=s_l$ from $n'$ samples. Such an error is just the variance of the underlying distribution of $s_l$, i.e., $ \mathbb{E}[(s_l-\mathbb{E}[s_l])^2]$, and it is straightforward to show that $n'S_{n'}^2$ forms an unbiased estimator. Similarly, one can show that $n'S_{n'}^2/n$ is an unbiased estimator from $n'$ samples for $\mathrm{Var}(\Bar{s}_n)$.

\captionsetup{labelformat = suppl}
\begin{figure}
    \centering
    \includegraphics[scale=0.75]{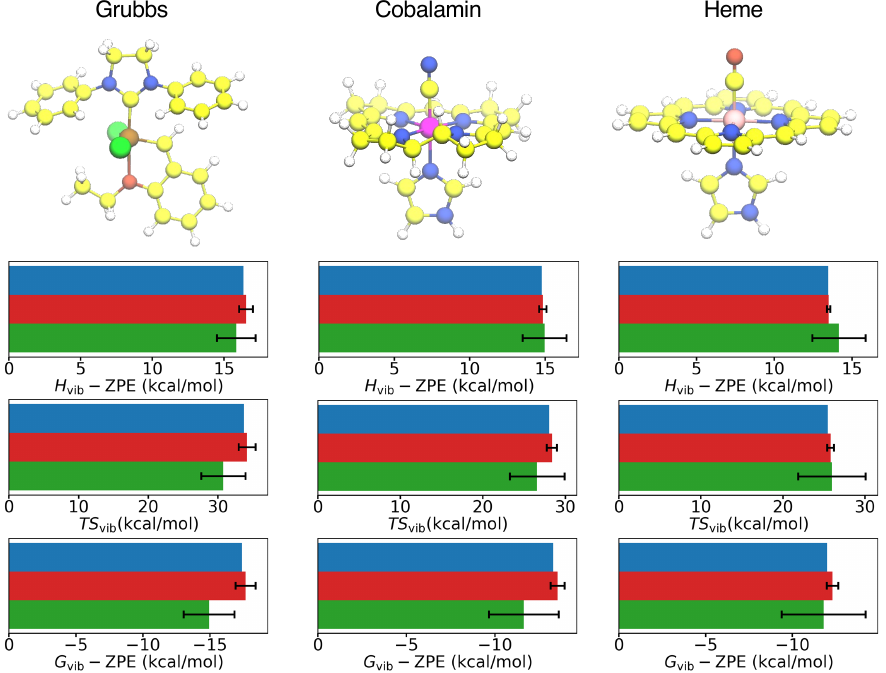}
    \caption{Thermal quantities at the DFT-level for a Grubbs catalyst, cobalamin, and heme. The blue bars indicate the exact DFT values, the red bars represent the average over 10 samples from correlated sampling using xTB as the low-level method, and the green bars are the average from 10 samples of direct sampling with DFT. The error bars are estimates for a 2-sample estimator computed from the total 10 samples. For these relatively small systems, 2 samples correspond to roughly 20\% of the cost of computing exact numerical Hessian.}
    \label{fig:metal}
\end{figure}

\section{Performance on transition-metal complexes}

Transition metal complexes provide an example of systems where empirical force fields are often inadequate and quantum mechanical energy functions are especially valuable. 
We considered three transition-metal complexes, namely a Grubbs catalyst, a cobalamin model, and a heme model, as challenging cases. The initial structure of the Grubbs catalyst was taken from Ref.~\cite{albalawi2021influence}, optimized with both xTB and B3LYP\cite{stephens1994ab} functional theory. The 6-311G basis set\cite{krishnan1980self,mclean1980contracted} was used for main group elements while the LanL2DZ basis\cite{hay1985ab} was used for the transition metal Ru. The stochastic sampling was performed using the same level of theory. The cobalamin structure was taken from Ref.~\cite{kornobis2011electronically}, and BP86\cite{becke1988density,perdew1986density}/6-31G(d)\cite{hehre1972self,hariharan1973influence,rassolov19986} was used for geometry optimization and sampling. The heme structure was taken from Ref.~\cite{harvey2000dft}, and we used the B3LYP/6-311G(d)\cite{krishnan1980self} level of theory. The DFT methods were chosen to be similar to the ones employed in the references from which the initial structures were obtained. In all the calculations, a Lanczos order of 16 was used, which we see to be sufficient to reproduce the exact results to within the small statistical error bars.

\red{
\section{Self-averaging in diamond nanocrystals}
In order to confirm that the observed small statistical error in large diamond crystals is the result of ``self-averaging'', instead of due to under-sampling of the larger coordinate space, we show in Supplementary Table~\ref{tab:selfave} the exact absolute free energy of the two diamond systems, computed from the exact xTB Hessian and from our method using 50 samples (numbers in kcal/mol). We see that the exact values all lie within one standard error of the sampled values, validating the error bars at both system sizes, despite the fixed number of samples as a function of system size. Moving from $\mathrm{C_{54}H_{54}}$ to $\mathrm{C_{128}H_{96}}$, the system size more than doubles, but the statistical error increases by much less (and in fact stays nearly constant), which is the self-averaging effect referred to in the text. } 

\begin{table}
    \centering
    \begin{tabular}{c|cccc}
        System & & ZPE & $\mathrm{H_{vib}-ZPE}$ & $\mathrm{TS_{vib}}$\\
        \hline
        $\mathrm{C_{54}H_{54}}$ &   Exact   & 599.8436 & 15.742 & 23.5286 \\
                                &   Sampled & 602$\pm$4 & 16.0$\pm$0.3& 24.0$\pm$0.6\\
        \hline
        $\mathrm{C_{128}H_{96}}$ &  Exact   & 1206.4788 & 31.9561 & 47.1107\\
                                 &  Sampled & 1209$\pm$5 & 32.0$\pm$0.4& 47.2$\pm$0.6\\
    \end{tabular}
    \caption{Absolute quantities of diamonds showing self-averaging.}
    \label{tab:selfave}
\end{table}

\red{
\section{Convergence with fitting order}
Here, we check the convergence with respect to the fitting order $m$. To obtain a basic understanding, we first examine the accuracy of Chebyshev polynomial fitting for the thermodynamic quantities, which can be assessed without using  any stochastic sampling. We do so by examining the accuracy of the Chebyshev expansion  over a range of frequencies (see Supplementary Figure~\ref{fig:cheby_mconv}). The Chebyshev error for a specific system can be estimated by averaging this fitting error over the frequency domain, weighted by the density of the system's vibrational states. 
The lowest vibrational frequency is 0 cm$^{-1}$, while the highest frequency of a chemical system is usually the bond vibration between a hydrogen and a heavy atom (typically $<$ 4000 cm$^{-1}$). Thus the only system dependence comes from the density of states. 
This is in contrast to the case of Chebyshev fitting in electronic structure, where the maximum frequency usually grows with system size. We see in both plots that the maximum pointwise deviation at $m=16$ is less than 1 kcal/mol over the range plotted (although we note that the entropy diverges at $0$ cm$^{-1}$ where the density of states also vanishes). 
The above convergence check cannot be directly carried out for stochastic Lanczos quadrature due to the need to specify some initial stochastic vector. We have therefore checked the convergence of the Lanczos method as a function of $m$ for several systems by  brute-force stochastic sampling, as shown in Supplementary Figures~\ref{fig:metal_mconv} and \ref{fig:tnks_mconv}. A Lanczos order of $m=16$ is generally seen to be sufficient to obtain a systematic error of 1 kcal/mol or less.
}

\begin{figure}
    \centering
    \includegraphics[scale=0.85]{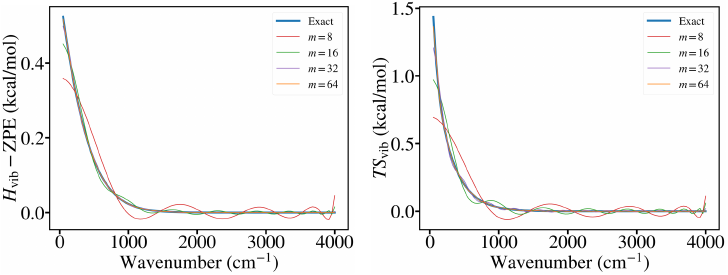}
    \caption{Chebyshev fitting to thermodynamic quantities at 298.15 K as a function of fitting order $m$. }
    \label{fig:cheby_mconv}
\end{figure}

\begin{figure}
    \centering
    \includegraphics[scale=0.75]{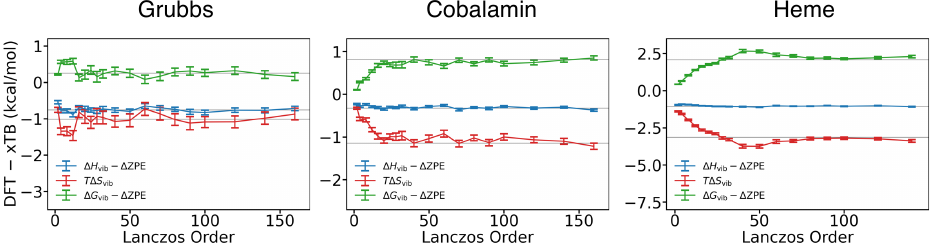}
    \caption{Convergence w.r.t. Lanczos order $m$ for transition metal complexes. Grey lines indicate the exact values. Stochastic errors are estimated from 500 samples.}
    \label{fig:metal_mconv}
\end{figure}

\begin{figure}
    \centering
    \includegraphics{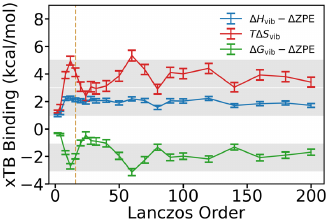}
    \caption{Convergence w.r.t. Lanczos order $m$ for TNKS ligand binding. Grey shaded area indicates $\pm$ 1 kcal/mol around the exact values. The vertical dashed line indicates $m=16$. Stochastic errors are estimated from 200 samples.}
    \label{fig:tnks_mconv}
\end{figure}

\bibliography{ref}